\documentclass[12pt]{article}
\usepackage[super,compress]{cite}
\usepackage{epsfig}
\usepackage{graphicx}
\usepackage{braket}

\newcommand{\be}{\begin{equation}}
\newcommand{\ee}{\end{equation}}
\newcommand{\bea}{\begin{eqnarray}}
\newcommand{\eea}{\end{eqnarray}}
\newcommand{\beq}{\begin{equation}}
\newcommand{\eeq}{\end{equation}}
\newcommand{\nn}{\nonumber}

\def\fun#1#2{\lower3.6pt\vbox{\baselineskip0pt\lineskip.9pt
\ialign{$\mathsurround=0pt#1\hfil##\hfil$\crcr#2\crcr\sim\crcr}}}

\begin{document}

\title{Non-strange and strange pentaquarks with hidden charm}

\author{
V.V. Anisovich$^+$, M.A. Matveev$^+$,
 J. Nyiri$^*$,
 A.V. Sarantsev$^{+ \diamondsuit}$,
A.N. Semenova$^+$ }

\maketitle

\begin{center}
{\it
$^+$National Research Centre ''Kurchatov Institute'':
Petersburg Nuclear Physics Institute, Gatchina, 188300, Russia}

{\it $^\diamondsuit$
Helmholtz-Institut f\"ur Strahlen- und Kernphysik,
Universit\"at Bonn, Germany}

{\it $^*$Institute for Particle and Nuclear Physics, Wigner RCP,
Budapest 1121, Hungary}

\end{center}

\begin{abstract}
Non-strange and strange pentaquaks with hidden charm are considered as
diquark-diquark-antiquark composite systems. Spin and isospin content of such exotic
states is discussed and masses are evaluated.

\end{abstract}



\section{Introduction}

The world of hadrons is a world of composite particles. The idea of quark structure of hadrons appeared long ago \cite{gell-mann}:
mesons are quark-antiquark systems, baryons are built of three quarks.
This structure provides a possibility for the
systematization of the hadrons and model calculation of the hadron characteristics.

For many decades this procedure worked perfectly, and it is still so for the mesons.
The mesonic sector of light quarks $(u,d,s)$ fits well
the flavours, the mesons are placed with a good accuracy on the Regge trajectories
and the trajectories of radial excitations  \cite{book3}.

However, the growing number of experimental results leads to
unexpected problems in the baryonic sector.
It turned out that much
less baryons were detected experimentally than the number of
predicted three-quark excitations.

Introduction of diquarks as pimary constituents for hadrons
can improve this contradictory situation (the notion of the diquark
was introdused also by
Gell-Mann \cite{gell-mann}).
If diquarks exist and can be considered as primary objects, the number of predicted
baryon states has to be smaller than in the three-quark scheme. In the framework of
such scheme of baryons \cite{qD} ''towers'' of baryonic states appear, i.e.
groups of resonances with the same masses but different spins. The existence of such
towers could give an explanation for the absence of signals coming from groups of
resonances which are shielded by their "neighbours".
Unfortunately, there are no experimental results demonstrating the
existence of such groups of states.
This may, however, not confirm
their absence since there was no real search for towers.

The scheme with constituent  quarks and diquarks suggests an existence
diquark-antidiquark mesons and diquark-diquark-antiquark baryons.

If the signal detected by the LHCb collaboration \cite{LHCbB} in the
spectrum $pJ/\psi$ of the reaction $\Lambda\to K^-(pJ/\psi)$ is
inducted by the pole singularity of the amplitude, this is a
definite confirmation of the fact that the baryon sector consists
not only of excitations of three-quark or quark-diquark degrees of
freedom, but also of a growing number of constituents.

Such a possibility for the growth of the hadron sector was
considered already long ago \cite{jaff,deru,stro}, and the vivid discussion
continues up to now for both the case of mesons
\cite{maiani,voloshin,ali,wein,brod} and baryons \cite{JW,ross}.

In a world of particles built of diquarks we arrived at a point where we
are able to define the quantum numbers of some new hadrons and estimate
the intervals between their probable masses. In the present paper we
try to do this on the basis of the systematization of mesons
\cite{ala-1} and baryons \cite{ala-2}.

In terms of the quark-diquark states the LHCb pentaquark
\cite{LHCbB} can be presented as a three-body system
$\bar c\cdot(cu)\cdot(ud)$ where $(cu)$ and $(ud)$ refer to heavy-light and
light-light diquarks. A diquark is a color antitriplet member, similar
to a antiquark, and the three-body system diquark-diquark-antiquark has
a color structure similar to that in low-lying baryons. It is
reasonable to suppose a similarity of color forces in three-quark
and diquark-diquark-antiquark systems.

The LHCb collaboration \cite{LHCbB} within study $\Lambda^0_b\to K^-J/\psi p$ decay
presents two candidates for pentaquarks decaying to the $p\,J/\psi$-channel.
 That are a
narrow peak  $5/2^?(4450\pm 4)$ with a width $\Gamma=(39\pm 24)$ MeV and a broad state
$3/2^?(4380\pm 38)$ with $\Gamma=(205\pm 94)$ MeV, the opposite parity of the states is
preferred. In Ref.~\citen{maia-polo} the variant of $3/2^-$ and $5/2^+$ is accepted, in
Ref.~\citen{ala-2} we consider a version with $3/2^+$ and $5/2^-$. Actually we suppose that the
narrow peak  $5/2^-(4450\pm 4)$ is a genuine S-wave pentaquark while
in the mass region $M_{pJ/\psi}\simeq 4380$ MeV other subruns can contribute thus
imitating the broad state.
That are final state rescatterings of outgoing hadrons and/or reflection of
possible exotic meson resonances
from the $K^-J/\psi$ channel. The matter is the S-wave diquark-antidiquark states with
open strangeness which have masses in the range of 4000-4550 MeV \cite{ala-1} and can be
produced in the $K^-J/\psi$ spectrum \cite{ala-3}.

The LHCb pentaquark states are now a subject of wide discussion
\cite{mikh,liu-wang,meis-olle,miro,he,lebe,meis}. In the present paper we continue
the discussion presenting possible classification of the non-strange and strange
pentaquarks with hidden charm and estimate their masses.

\section{Pentaquarks as diquark-diquark-antiquark systems }

We discuss a scheme in which the exotic states are formed by standard
QCD-motivated interactions (gluonic exchanges, confinement forces) but with diquarks
as constituents.
In the color space we write for the pentaquark:
\bea
&&
P^+_{\bar ccuud }=
\epsilon_{\alpha\beta\gamma}\bar c^\alpha(cu)^\beta(ud)^\gamma\,+\,
{\rm permutations\, of\, the}\, u,d\,{\rm quarks }
\quad,
\\
&&
(cu)^\beta=\epsilon^{\beta\beta'\gamma'}c_{\beta'}u_{\gamma'},
\quad
(ud)^\gamma=\epsilon^{\gamma\beta''\gamma''}u_{\beta''}d_{\gamma''},
\nn
\eea
where $\alpha,\beta,\gamma$ refer to color indices.

\subsubsection{Spin structure of the pentaquarks}

We work with two diquarks, scalar $S(0^+)$ one and axial-vector $A(1^+)$ one. In terms of these
diquarks the color-flavor wave function of pentaquark reads:
\be \label{2}
P_{\bar c\cdot
(cq)(q'q'')}=\bar c^\alpha\cdot \epsilon_{\alpha\beta\gamma}\,
\begin{tabular}{|l|}
$S^\beta_{(cq)}$\\
$A^\beta_{(cq)}$
\end{tabular}
\cdot
\begin{tabular}{|l|}
$S^\gamma_{(q'q'')}$\\
$A^\gamma_{(q'q'')}$
\end{tabular},   \qquad  \mbox{ with }
q,q',q''=u,d,s.
\ee
It results in six diquark-diquark states:
\be          \label{3}
P_{\bar c\cdot (cq)(q'q'')}= \bar c^\alpha\cdot
\begin{tabular}{|l|}
$(S_{(cq)}S_{(q'q'')})^\alpha(0^{+})$\\
$(S_{(cq)}A_{(q'q'')})^\alpha(1^{+})$\\
$(A_{(cq)}S_{(q'q'')})^\alpha(1^{+})$\\
$(A_{(cq)}A_{(q'q'')})^\alpha(0^{+})$\\
$(A_{(cq)}A_{(q'q'')})^\alpha(1^{+})$\\
$(A_{(cq)}A_{(q'q'')})^\alpha(2^{+})$
\end{tabular}
\ee
with the spin-parity numbers for diquark-diquark subsystems
equal to $J^P= 0^+, 1^+,2^+$.

\subsubsection{Isospin structure of the pentaquarks}

We have the following isospin-spin structure for the
diquarks:
\bea         \label{4}
&&
S_{(cq)}(I=1/2,\,J=0),\qquad A_{(cq)}(I=1/2,\,J=1),
 \\ &&
S_{(q'q'')}(I=0,\,J=0),\qquad A_{(q'q'')}(I=1,\,J=1),
\nn
\eea
and the isospin-spin sector of the pentaquarks $P(I,J^P)$ reads:
\be \label{5}
P(I,J^P)=\begin{tabular}{l}
$\bar c S_{(cq)}S_{(q'q'')}(0^+):\quad$$P(\frac12,\frac12^{-})$,\\
$\bar c S_{(cq)}A_{(q'q'')}(1^+):\quad$$P(\frac12,\frac12^{-})$,
$ P(\frac12,\frac32^{-})$, $P(\frac32,\frac12^{-})$, $P(\frac32,\frac32^{-})$, \\
$\bar c A_{(cq)}S_{(q'q'')}(1^+):\quad$$P(\frac12,\frac12^{-})$, $P(\frac12,\frac32^{-})$,\\
$\bar c A_{(cq)}A_{(q'q'')}(0^+):\quad$$P(\frac12,\frac12^{-})$, $P(\frac32,\frac
12^{-})$,\\
$\bar c A_{(cq)}A_{(q'q'')}(1^+):\quad$$P(\frac12,\frac12^{-})$,
$P(\frac12,\frac32^{-})$, $P(\frac32,\frac12^{-})$, $P(\frac32,\frac32^{-})$, \\
$\bar c A_{(cq)}A_{(q'q'')}(2^+):\quad$$P(\frac12,\frac32^{-})$,
$P(\frac12,\frac52^{-})$, $P(\frac32,\frac32^{-})$, $P(\frac32,\frac52^{-})$.
\end{tabular}
 \ee

\subsubsection{ Diquark-antidiquark mesons: masses  and spin splitting
parameter}

In Ref.~\citen{ala-1} the diquark-antidiquark
mesons, $(cq)\cdot (\bar c\bar q')$, were studied, we use this study as a guide
for consideration of the pantaquarks.

The masses of the heavy diquarks were determined in Ref.~\citen{ala-1},
the masses of the light ones were determined in the previous studies \cite{qD}. The masses read:
\bea       \label{6}
&&
\begin{tabular}{ll}
$m_{S(q'q'')}=650\pm 50 \mbox{ MeV}$\,, & $m_{A(q'q'')}=750\pm 50 \mbox{ MeV}$\,, \\
$m_{S(sq')}=770\pm 50 \mbox{ MeV}$\,,   & $m_{A(sq')}=870\pm 50 \mbox{ MeV}$\,, \\
$m_{S(ss)}=900\pm 50 \mbox{ MeV}$\,,    & $m_{A(ss)}=1000\pm 50 \mbox{ MeV}$\,,  \\
$m_{S(cq)}=2000\pm 50 \mbox{ MeV}$\,,   & $m_{A(cq)}=2050\pm 50 \mbox{ MeV}$\,, \\
$m_{S(cs)}=2100\pm 50 \mbox{ MeV}$\,,   & $m_{A(cs)}=2150\pm 50 \mbox{ MeV}$\, .
\end{tabular}
\eea
For a rough estimation of the tetraquark masses the mass-splitting formula
was written in Ref.~\citen{ala-1} as:
\be  \label{7}
M^{(J)}_{(cq)\cdot (\bar c\bar q')}=m_{(cq)}+m_{(\bar c\bar q')}+
J(J+1)\,\Delta
\,.
\ee
The value $\Delta=70\pm 10$ MeV fits the data.
Within errobars the value for spin splitting parameter $\Delta$ coincides with that for
$c\bar c$ mesons. Indeed, for $J/\psi$ and $\eta_c$ one has
$m_{J/\psi}-m_{\eta_{c}}\simeq 120$ MeV that gives $\Delta_{c\bar c}\simeq 60$ MeV.
 For estimation of masses of strange diquark-antidiquark states
the same parameter $\Delta$ is used in Ref.~\citen{ala-1}.

\subsection{Estimation of masses of the pentaquarks}

It was understood relatively long ago \cite {ZS,RGG,glashow}
that the mass splitting of hadrons can be well described in the framework of the quark
model by the short-ranged spin-spin interactions of the constituents. For mesons and
baryons the mass formulae discussed by Glashow \cite {glashow} read:
\bea \label{8}
&&
M_M=\sum\limits_{j=1,2}m_{q(j)}+a
\frac{\vec{s}_1\vec{s}_2}{m_{q(1)}m_{q(2)}},
\\
&&
M_B=\sum\limits_{j=1,2,3}m_{q(j)}+b
\sum\limits_{j>\ell}\frac{\vec{s}_j\vec{s}_\ell}{m_{q(j)}m_{q(\ell)}},
\nn
\eea
where  $\vec{s}_j$ and  $m_{q(j)}$
refer to spins and masses of the constituents.
Mass splitting parameters in (\ref{8}), $a$ and $b$,
are characterized by  a size of the color-magnetic interaction in the discussed
hadron, the short-range interaction is supposed in Ref.~\citen{glashow}.
For the 36-plet mesons and 56-plet baryons formulae of Eq.~(\ref{8}) work well.
The only exception is the pion, its calculated mass is $\sim$350 MeV that point out
an existence of additional forces in the pseudoscalar channel (possibly, istanton-induces
forces, see Ref.~\citen{shuryak}).

Operating with mass-splitting term $J(J+1)\Delta$ we have
for the light hadrons:
$\Delta_{\Delta-N} =67\; {\rm MeV}$, $\Delta_{\omega-\eta} =115\; {\rm MeV}$,
and for charmonium mesons: $\Delta_{J/\psi-\eta_c} =60\; {\rm MeV}$.
For the tetraquark states with hidden charm and strangeness we obtain \cite{ala-1}:
$\Delta=70\pm 10$ MeV.

For evaluation of pentaquark masses we use similar procedure.
First, let us estimate the masses of diquark-diquark subsystems $(cq)\cdot (q'q'')$.
For the diquark-diquark subsystem we write:
\be  \label{9}
M_{(cq)\cdot (q'q'')}=m_{(cq)}+m_{(q'q'')}+
J_{(cq)\cdot (q'q'')}\bigg(J_{(cq)\cdot (q'q'')}+1\bigg)\,\Delta_{(cq)\cdot (q'q'')}
\,,
\ee
with the splitting parameter being of the order of that for the
diquark-antidiquark system, $\Delta_{(cq)\cdot (q'q'')}\sim 70$ MeV.
As the next step
we write masses of
the pentaquarks as a sum of masses of the constituents plus spin splitting term:
\bea   \label{10}
M_{\bar c\cdot (cq)(q'q'')}&=& m_{\bar c}+M_{(cq)\cdot (q'q'')}+
J_{\bar c\cdot (cq)(q'q'')}\bigg(J_{\bar c\cdot (cq)(q'q'')}+1\bigg)\,
\Delta_{\bar c\cdot (cq)(q'q'')}\\
&\simeq &m_{\bar c}+M_{(cq)\cdot (q'q'')}
,
\nn
\eea
where the mass of the constituent antiquark $\bar c$ is equal to
$m_c=1300\pm 50\mbox { MeV}$ \cite{ADMNS-cc,mano}.
In the last line of Eq.~(\ref{10}) we neglect the mass splitting term,
$\Delta_{\bar c\cdot (cq)\cdot (q'q'')}\to 0$. The experimental data \cite{LHCbB} require
this parameter to be small, $\Delta_{\bar c\cdot (cq)\cdot (q'q'')}< 15$ MeV.
We put it zero in order to manage without cumbersome terms.

Following the Eq.~(\ref{10}) we write for the low-laying
non-strange pentaquarks $P_{\bar c\cdot (cq)(q'q'')}$ with $q,q',q''=u,d$:
\be    \label{11}
\begin{tabular}{|l|l|}
$P_{\bar c\cdot (cq)(q'q'')},\qquad I=\frac12$&  $P_{\bar c\cdot (cq)(q'q'')},\qquad I=\frac32 $     \\
\hline
$P_{\bar cS_{cq}S_{q'q''}}^{(\frac12,\frac12^{-})}(3800)$&\\
$P_{\bar cS_{cq}A_{q'q''}}^{(\frac12,\frac12^{-})}(4190)$,
$P_{\bar cS_{cq}A_{q'q''}}^{(\frac12,\frac32^{-})}(4190)$ &
$P_{\bar cS_{cq}A_{q'q''}}^{(\frac32,\frac12^{-})}(4190)$,
$P_{\bar cS_{cq}A_{q'q''}}^{(\frac32,\frac32^{-})}(4190)$ \\
$P_{\bar cA_{cq}S_{q'q''}}^{(\frac12,\frac12^{-})}(4140)$,
$P_{\bar cA_{cq}S_{q'q''}}^{(\frac12,\frac32^{-})}(4140)$&\\
$P_{\bar cA_{cq}A_{q'q''}}^{(\frac12,\frac12^{-})}(4100)$&
$P_{\bar cA_{cq}A_{q'q''}}^{(\frac32,\frac12^{-})}(4100)$\\
$P_{\bar cA_{cq}A_{q'q''}}^{(\frac12,\frac12^{-})}(4240)$,
$P_{\bar cA_{cq}A_{q'q''}}^{(\frac12,\frac32^{-})}(4240)$&
$P_{\bar cA_{cq}A_{q'q''}}^{(\frac32,\frac12^{-})}(4240)$,
$P_{\bar cA_{cq}A_{q'q''}}^{(\frac32,\frac32^{-})}(4240)$ \\
$P_{\bar cA_{cq}A_{q'q''}}^{(\frac12,\frac32^{-})}(4520)$,
$P_{\bar cA_{cq}A_{q'q''}}^{(\frac12,\frac52^{-})}(4520)$ &
$P_{\bar cA_{cq}A_{q'q''}}^{(\frac32,\frac32^{-})}(4520)$,
$P_{\bar cA_{cq}A_{q'q''}}^{(\frac32,\frac52^{-})}(4520)$
\end{tabular}
 \ee
Masses are given in MeV units, the uncertainty in the determination of masses is of the order $\pm$150 MeV.
We consider the state
$P_{A_{(cq)}A_{(q'q'')}}^{(\frac12,\frac52^{-})}(4520\pm 150)$
as pentaquark seen in Ref.~\citen{LHCbB} which reveals itself as a narrow peak
at $M_{pJ/\psi}$=4450 MeV in the ($pJ/\psi$)-spectrum.

\subsection{Pentaquarks with open strangeness}

We have two sets of pentaquark states $P_{\bar c\cdot (cq)(sq')}$ with open strangeness $S=1$ :
 \be    \label{12}
\begin{tabular}{|l|l|}
$P_{\bar c\cdot (cq)(sq')},\qquad I=0$& $P_{\bar c\cdot (cq)(sq')},\qquad I=1 $\\
\hline
$P_{\bar c S_{cq}S_{sq'}}^{(0,\frac12^{-})}(4070)$
&
$P_{\bar c S_{cq}S_{sq'}}^{(1,\frac12^{-})}(4070)$
\\
$P_{\bar c S_{cq}A_{sq'}}^{(0,\frac12^{-})}(4310)$,
$P_{\bar c S_{cq}A_{sq'}}^{(0,\frac32^{-})}(4310)$&
$P_{\bar c S_{cq}A_{sq'}}^{(1,\frac12^{-})}(4310)$,
$P_{\bar c S_{cq}A_{sq'}}^{(1,\frac32^{-})}(4310)$
\\
$P_{\bar c A_{cq}S_{sq'}}^{(0,\frac12^{-})}(4260)$,
$P_{\bar c A_{cq}S_{sq'}}^{(0,\frac32^{-})}(4260)$&
$P_{\bar c A_{cq}S_{sq'}}^{(1,\frac12^{-})}(4260)$,
$P_{\bar c A_{cq}S_{sq'}}^{(1,\frac32^{-})}(4260)$
\\
$P_{\bar c A_{cq}A_{sq'}}^{(0,\frac12^{-})}(4220)$&
$P_{\bar c A_{cq}A_{sq'}}^{(1,\frac12^{-})}(4220)$
\\
$P_{\bar c A_{cq}A_{sq'}}^{(0,\frac12^{-})}(4360)$,
$P_{\bar c A_{cq}A_{sq'}}^{(0,\frac32^{-})}(4360)$&
$P_{\bar c A_{cq}A_{sq'}}^{(1,\frac12^{-})}(4360)$,
$P_{\bar c A_{cq}A_{sq'}}^{(1,\frac32^{-})}(4360)$
\\
$P_{\bar c A_{cq}A_{sq'}}^{(0,\frac32^{-})}(4640)$,
$P_{\bar c A_{cq}A_{sq'}}^{(0,\frac52^{-})}(4640)$&
$P_{\bar c A_{cq}A_{sq'}}^{(1,\frac32^{-})}(4640)$,
$P_{\bar c A_{cq}A_{sq'}}^{(1,\frac52^{-})}(4640)$
\end{tabular}
 \ee
and $P_{\bar c\cdot (cs)(qq')}$ with $S=1$:
 \be    \label{13}
\begin{tabular}{|l|l|}
$P_{\bar c\cdot (cs)(qq')},\qquad I=0$& $P_{\bar c\cdot (cs)(qq')},\qquad I=1 $ \\
\hline
$P_{\bar c S_{cs}S_{qq'}}^{(0,\frac12^{-})}(4050)$
&
\\
&
$P_{\bar c S_{cs }A_{qq'}}^{(1,\frac12^{-})}(4290)$,
$P_{\bar c S_{cs}A_{qq'}}^{(1,\frac32^{-})}(4290)$
\\
$P_{\bar c A_{cs}S_{qq'}}^{(0,\frac12^{-})}(4240)$,
$P_{\bar c A_{cs}S_{qq'}}^{(0,\frac32^{-})}(4240)$&
\\
&
$P_{\bar c A_{cs}A_{qq'}}^{(1,\frac12^{-})}(4200)$
\\
&
$P_{\bar c A_{cs}A_{qq'}}^{(1,\frac12^{-})}(4340)$,
$P_{\bar c A_{cs}A_{qq'}}^{(1,\frac32^{-})}(4340)$
\\
&
$P_{\bar c A_{cs}A_{qq'}}^{(1,\frac32^{-})}(4620)$,
$P_{\bar c A_{cs}A_{qq'}}^{(1,\frac52^{-})}(4620)$
\end{tabular}
 \ee
In this sector we see practically overlapping states with equal
$(I,J)$ values, like, e.g.,\linebreak
${P_{\bar c S_{(cq)}S_{(sq')}}^{(0,\frac12^{-})}(4070\pm150)}$
and
$P_{\bar c S_{(cs)}S_{(qq')}}^{(0,\frac12^{-})}(4050\pm150)$.
Such overlapping states can be intensely intermix
so that one of the states accumulates the width of the other ones
\cite{PR-exotic} - as a result we see a narrow peak on a
broad substrate.

A set of the pentaquark states $P_{\bar c\cdot (cq)(ss)}$ and $P_{\bar c\cdot (cs)(sq)}$ with strangeness $S=2$ can be written as:
 \be    \label{14}
\begin{tabular}{|l|l|}
$P_{\bar c\cdot (cq)(ss)},\qquad I=\frac 12$& $P_{\bar c\cdot (cs)(sq)},\qquad I=\frac 12 $
\\
\hline
$P_{\bar c S_{cq}S_{ss}}^{(\frac12,\frac12^{-})}(4200)$&
$P_{\bar c S_{cs}S_{sq}}^{(\frac12,\frac12^{-})}(4170)$
\\
$P_{\bar c S_{cq}A_{ss}}^{(\frac12,\frac12^{-})}(4440)$,
$P_{\bar c S_{cq}A_{ss}}^{(\frac12,\frac32^{-})}(4440)$&
$P_{\bar c S_{cs}A_{sq}}^{(\frac12,\frac12^{-})}(4410)$,
$P_{\bar c S_{cs}A_{sq}}^{(\frac12,\frac32^{-})}(4410)$
\\
$P_{\bar c A_{cq}S_{ss}}^{(\frac12,\frac12^{-})}(4390)$,
$P_{\bar c A_{cq}S_{ss}}^{(\frac12,\frac32^{-})}(4390)$&
$P_{\bar c A_{cs}S_{sq}}^{(\frac12,\frac12^{-})}(4460)$,
$P_{\bar c A_{cs}S_{sq}}^{(\frac12,\frac32^{-})}(4460)$
\\
$P_{\bar c A_{cq}A_{ss}}^{(\frac12,\frac12^{-})}(4350)$&
$P_{\bar c A_{cs}A_{sq}}^{(\frac12,\frac12^{-})}(4320)$
\\
$P_{\bar c A_{cq}A_{ss}}^{(\frac12,\frac12^{-})}(4490)$,
$P_{\bar c A_{cq}A_{ss}}^{(\frac12,\frac32^{-})}(4490)$&
$P_{\bar c A_{cs}A_{sq}}^{(\frac12,\frac12^{-})}(4460)$,
$P_{\bar c A_{cs}A_{sq}}^{(\frac12,\frac32^{-})}(4460)$
\\
$P_{\bar c A_{cq}A_{ss}}^{(\frac12,\frac32^{-})}(4770)$,
$P_{\bar c A_{cq}A_{ss}}^{(\frac12,\frac52^{-})}(4770)$&
$P_{\bar c A_{cs}A_{sq}}^{(\frac12,\frac32^{-})}(4740)$,
$P_{\bar c A_{cs}A_{sq}}^{(\frac12,\frac52^{-})}(4740)$
\end{tabular}
 \ee
In this sector we also see nearly overlapping states.

\subsection{The LHCb data: where are other exotic states?}

The LHCb peak $5/2^?(4450\pm 4)$ being interpreted as pentaquark state
arises questions.
Existence of pentaquark means an existence of an  assembly of  pentaquarks.
Specifically, the pentaquark with $J^P=5/2^-$ should be accompanied by states with
smaller masses and $J^P=3/2^-,1/2^-$, see Eq.~(\ref{11}); the similar  assembly
exists for positive parity pentaquarks. The multiplet neighbours are not observed in
Ref.~\citen{LHCbB}. Special studies are needed to clarify if these  assemblies of states do realized or not.

Moreover, if pentaquark states exist the tetraquark states exist as well. It means that
in the $KJ/\psi$ sector diquark-antidiquark states
$(cs)\cdot (\bar c \bar d)$ can be observed in the mass region 4000-4500 MeV,
production of the low-lying tetraquarks is not forbidden in the decay
$\Lambda^0_b\to K^-J/\psi p$.
But definite indication for production of such states is absent in Ref.~\citen{LHCbB}.
Production of the tetraquark states $(cs)\cdot (\bar c \bar d)$ in
the decay $\Lambda^0_b\to K^-J/\psi p$ would mean crossing of tetraquark and petraquark
signals on Dalitz-plot thus complicating interpretation of the data.

\section{Conclusion}

The scheme of the low-lying S-wave pentaquarks is developed
on the basis of the study of the tetraquark states \cite{ala-1}
and results of the LHCb-collaboration \cite{LHCbB}.
Following Ref.~\citen{ala-2} we suggest that the narrow peak $5/2^?(4450\pm 4)$ with
$\Gamma=(39\pm 24)$ MeV presents the
$S$-wave pentaquark denoted here as
$P_{S_{(cq)}A_{(q's)}}^{(\frac12,\frac52^{-})}(4520\pm 150)$, see Eq.~(\ref{11}). Than the broad bump,  $3/2^?(4380\pm 38)$ with effective width
$\Gamma=(205\pm 94)$~MeV, also seen in the $(p\,J/\psi)$
spectrum, can be related with possible
resonances of the $K^-J/\psi$-channel located in the mass region 4000-4500 MeV.

According to the diquark-diquark-antiquark scheme the mass interval 4040-4500 MeV
contains several exotic baryon resonances with  $J^P=3/2^-,1/2^-$. Interplay
of these baryon states with exotic mesons from the $K^-J/\psi$-channel
may be crucial for decipher the Dalitz-plot.

Search for exotic baryon resonances in other channals may be effective, for example,
decay of the low-laying $1/2^-$-states into the $(p\,\eta_c)$-channel looks as dominant.

\section*{Acknowledgments}

The work was supported by grant RSCF-14-22-00281.

\end{document}